\documentstyle[aps]{revtex}

\begin{document}

\section{Introduction}

The propagation of electromagnetic waves through plasmas has been
extensively studied from a classical viewpoint. This is usually justified in
most cases by the physical conditions regarding the plasma temperature and
density at least when the considered wavelengths are not too short.

In the classical approach the plasma operates as a sort of active filter
both absorbing and distorting electromagnetic waves in various ways
depending on the plasma parameters, its homogeneity, the presence of an
external magnetic field and of course the frequency of the wave. To have an
idea of this complicated methodology and of the kind of results one obtains
it is useful to have a look to ref.\cite{ginzburg1}. It is remarkable that,
according to the different conditions, various frequency shifts arise.

Another classical approach, particularly fit for optical, or shorter,
wavelengths is that which considers the interaction of the electromagnetic
wave with a fluctuating medium as a scattering by inhomogeneities inside it.
This analysis has been made in refs.\cite{wolf1}, \cite{wolf2}, \cite{wolf3}
and leads again to frequency shifts of the incoming radiation depending now
also on the scattering angle. In particular a blue shift is found for
forward scattering, as a manifestation of the Rayleigh scattering.

In principle however the quantum aspects of the interaction of the wave with
the plasma should not be overlooked. This is likely to be especially true in
some situations of astrophysical or even cosmological interest, where the
density of the plasma is rather low and the distances are so high to allow
also small effects to pile up.

In this paper we shall precisely investigate the quantum effects using
standard quantum electrodynamics and a perturbative treatment, as outlined
in section \ref{II}. The method will produce also the known classical
features of the propagation of electromagnetic waves through a plasma, as we
show in section \ref{IV}. On the quantum side, we take into account in
section \ref{III} the many body nature of a plasma that entrains the
appropriate fermionic statistic.

In the non relativistic limit we find, for a low density locally homogeneous
(i.e. homogeneous on the scale of a few wavelengths) plasma, a blue shift of
the photon frequency (this has the same sign as that of the classical
results in \cite{wolf1}, \cite{wolf2}, \cite{wolf3}, though now the effect
has a completely different origin).

Finally in section \ref{V} we discuss the validity conditions of our
approximations, in particular those concerning the possibility of
overlooking relativistic corrections.

Just to fix ideas and exemplify our low density plasma, we refer, in various
parts of the paper, to numerical values of the parameters of the order of
those valid for the solar corona, i.e. a number density $\sim 10^6$
electrons/cm$^3$ and a temperature $\sim 10^6$ K.

\section{Basic assumptions and outline of the method}

\label{II}We consider a situation where an electromagnetic wave, plane and
monochromatic, coexists with a plasma of electrons with a numerical density
distribution $n({\bf x})$. The propagation is along the $x$ axis.

The ''unperturbed'' state is obtained when the coupling between the wave and
the plasma is set to zero; consequently the energy distribution of the
electrons is that of a fermionic plasma of temperature $T$ and density $n$
restrained into a potential well and the wave has a frequency $\omega $.

Let us now switch on the coupling and, assuming the interaction energy to be
small, we determine the shift in the energies as a perturbation of the
''initial'' (i.e. uncoupled) situation. Supposing that the interaction is
set up in a finite time lapse and comparing the situations at $-\infty $ in
time with that at $+\infty $, we mimic the actual process of an incoming
plane wave of frequency $\omega $ and an outgoing one of frequency $\omega
+\delta \omega $.

The technique actually used to compute the shift in the frequency of the
wave is that of the time independent perturbations. The interaction
Hamiltonian is: 
\begin{equation}
H_I=\text{ }-\frac ec\int {\bf j}\cdot {\bf A}\,\,d{\bf r}  \label{a1}
\end{equation}
where ${\bf A}$ is the vector potential operator of the wave and ${\bf j}$
is the current density operator appropriate for this problem; the volume
integral is limited to the confinement region of the plasma. The complete
non relativistic expression for ${\bf j}$ when an electromagnetic
interaction is present is \cite{Landau}: 
\begin{equation}
{\bf j}=-\frac{i\hbar }{2m}\left( \Phi ^{\dagger }{\bf \nabla }\Phi -\Phi 
{\bf \nabla }\Phi ^{\dagger }\right) -\frac e{mc}\Phi ^{\dagger }\Phi {\bf A}
\label{densita}
\end{equation}

Cast in the second quantization formalism, the interaction Hamiltonian is%
\cite{sakurai}: 
\begin{equation}
H_I=H_I^{^{\prime }}+H_I^{^{\prime \prime }}  \label{a0}
\end{equation}
\begin{equation}
H_I^{\prime }:=\left( \sum_{{\bf k}}b_{{\bf k}}\otimes \sum_{{\bf q},{\bf q}%
^{^{\prime }}}g_{{\bf q},{\bf q}^{^{\prime }}}^{{\bf k}}a_{{\bf q}}^{\dagger
}a_{{\bf q}^{^{\prime }}}+h.c.\right)  \label{a5}
\end{equation}
\begin{equation}
H_I^{\prime \prime }:=\frac{e^2}{2mc^2}\frac \hbar V\sum_{{\bf k,k}^{\prime
},{\bf q},{\bf q}^{^{\prime }}}\frac{\delta _{{\bf k-k}^{\prime },{\bf q-q}%
^{\prime }}}{\sqrt{\omega \omega ^{\prime }}}b_{{\bf k}^{\prime }}^{\dagger
}b_{{\bf k}}\otimes a_{{\bf q}^{\prime }}^{\dagger }a_{{\bf q}}  \label{a6}
\end{equation}
where we defined 
\begin{equation}
g_{{\bf q},{\bf q}^{^{\prime }}}^{{\bf k}}=\frac em\sqrt{\frac{\hbar ^3}{%
2V\omega }}{\bf q}^{\prime }\cdot {\bf u}_{{\bf k}}\delta _{{\bf k},{\bf q}%
^{\prime }{\bf -q}}  \label{a7}
\end{equation}

The $b^{\dagger }$'s and the $b$'s are the bosonic creation and annihilation
operators associated to the photons; the $a$'s are fermionic (electronic)
operators. The energy of the photon is $\hbar \omega ,$ its momentum is $%
\hbar {\bf k}$; of course $\omega =ck$; the polarization of the photon is
expressed by the unitary vector ${\bf u}_{{\bf k}}$; ${\bf k}$'s and ${\bf q}
$'s are respectively photonic and electronic wave vectors; primes denote
intermediate state variables; finally, the unessential spin and polarization
indices have been dropped, though an average over them in the final formulae
is performed.

The last term in (\ref{densita}) is usually omitted under the explicit, and
some times implicit, assumption it gives contributions small with respect to
those coming from the first.

The frequency shift of the external photon can be written in the form 
\[
\delta \omega =\delta \omega ^{\prime }+\delta \omega ^{\prime \prime } 
\]
where $\delta \omega ^{\prime }$ and $\delta \omega ^{\prime \prime }$ are
respectively the contributions due to $H_I^{\prime }$ and $H_I^{\prime
\prime }$. In this paper we calculate $\delta \omega $ for a low density
plasma, with a perturbative treatment up to order $\alpha =e^2/\hbar c$
(this is equivalent to keep the first order term in $H_I^{\prime \prime }$
and the second in $H_I^{\prime }$).

\section{The quantum frequency shift}

\label{III}We first consider the contributions to the shift due to $%
H_I^{\prime }$.

Let $H_{em}$ and $H_P$ be the electromagnetic field and the plasma free
Hamiltonians; ${\cal H}_{em}$ and ${\cal H}_P$ are respectively the bosonic
Fock space associated to the photons and the fermionic Fock space associated
to the electrons; $\Phi _{{\bf k}}\in $ ${\cal H}_{em}$ and $\Psi _{{\bf q}%
}\in $ ${\cal H}_P$ are the one particle electronic and photonic wave
functions.

We calculate the shift $\Delta E_{\omega ,{\bf q}}$ of the value of the
energy of the state $\Phi _{{\bf k}}\otimes \Psi _{{\bf q}}$ of ${\cal H}%
_{em}\otimes {\cal H}_P$. Since $\Phi _{{\bf k}}\otimes \Psi _{{\bf q}}$ is
an eigenvector of $H_{em}+H_P$ we can use the second order perturbation
theory (the first order energy shift is zero). The energy shift of the state 
$i$ is 
\begin{equation}
\Delta E_i=\sum_{j\neq i}\frac{\left( H_I\right) _{ij}\left( H_I\right) _{ji}%
}{E_i-E_j}\left( 1-\nu _j\right)  \label{b1}
\end{equation}
where $j$ labels any (normalized) eigenvector of $H_{em}+H_P$ and $\nu _j$
is the probability that the state $j$ is occupied by an electron.

The only non vanishing contributions are those with respect to $j$ states of
the form $j_1:=\Phi _{{\bf 0}}\otimes \Psi _{{\bf q^{\prime }}}$ , $%
j_2:=\left( \frac 1{\sqrt{2}}\Phi _{{\bf k}}\otimes \Phi _{{\bf k}^{\prime
}}\right) \otimes \Psi _{{\bf q^{\prime }}}$ or $j_3:=\left( \frac 1{\sqrt{2}%
}\Phi _{{\bf k}^{\prime }}\otimes \Phi _{{\bf k}}\right) \otimes \Psi _{{\bf %
q^{\prime }}}$, ($\Phi _{{\bf 0}}$ is the electromagnetic vacuum). It is
easy to show that 
\[
\left( H_I^{\prime }\right) _{j_1i}=g_{{\bf q},{\bf q^{\prime }}}^{{\bf k}} 
\]
\[
\left( H_I^{\prime }\right) _{j_2i}=\left( H_I^{\prime }\right) _{j_3i}=%
\frac 1{\sqrt{2}}g_{{\bf q^{\prime }},{\bf q}}^{{\bf k}^{\prime }} 
\]
So the energy shift induced by $H_I^{\prime }$ on $\Phi _{{\bf k}}\otimes
\Psi _{{\bf q}}$ is \cite{sissa}

\begin{equation}
\Delta E_{\omega ,{\bf q}}^{\prime }=\Delta E_{\omega ,{\bf q}}^{\left(
1\right) }+\Delta E_{\omega ,{\bf q}}^{\left( 2\right) }  \label{b2}
\end{equation}
where 
\begin{equation}
\Delta E_{\omega ,{\bf q}}^{\left( 1\right) }:=\sum_{{\bf q}^{^{\prime
}}}\left( 1-\nu _{q^{^{\prime }}}\right) \left[ \frac{\left| g_{{\bf q},{\bf %
q}^{\prime }}^{{\bf k}}\right| ^2}{\hbar \omega +\epsilon _q-\epsilon
_{q^{\prime }}}-\frac{\left| g_{{\bf q^{\prime }},{\bf q}}^{{\bf k}}\right|
^2}{\hbar \omega +\epsilon _{q^{\prime }}-\epsilon _q}\right]  \label{b3}
\end{equation}
\begin{equation}
\Delta E_{\omega ,{\bf q}}^{\left( 2\right) }:=\sum_{{\bf k}^{\prime }}\sum_{%
{\bf q}^{\prime }}\left( 1-\nu _{q^{\prime }}\right) \frac{\left| g_{{\bf %
q^{\prime }},{\bf q}}^{{\bf k}^{\prime }}\right| ^2}{\epsilon _q-\epsilon
_{q^{\prime }}-\hbar \omega ^{\prime }}  \label{b4}
\end{equation}
where, for a Fermi gas, in low density approximation: 
\begin{equation}
\nu _q=\frac 1{\frac 1ze^{\beta \epsilon _q}+1}\simeq ze^{-\beta \epsilon _q}
\label{b5}
\end{equation}
In order to obtain the energy shift, we have to take the mean value of $%
\Delta E_{\omega ,{\bf q}}^{\prime }$ with respect to the possible states of
the electrons $\Psi _{{\bf q}}.$ We have 
\[
\Delta E_\omega ^{\prime }=\Delta E_\omega ^{\left( 1\right) }+\Delta
E_\omega ^{\left( 2\right) } 
\]
\begin{equation}
\Delta E_\omega ^{\left( 1\right) }:=\sum_{{\bf q}}\nu _{q\,}\Delta
E_{\omega ,{\bf q}}^{\left( 1\right) }  \label{b6}
\end{equation}
\begin{equation}
\Delta E_\omega ^{\left( 2\right) }:=\sum_{{\bf q}}\nu _{q\,}\Delta
E_{\omega ,{\bf q}}^{\left( 2\right) }  \label{b66}
\end{equation}

We shall consider the contribution to the shift $\Delta E_\omega ^{\left(
1\right) }$ in the next section.

The term $\Delta E_\omega ^{\left( 2\right) }$ gives a divergent
contribution \cite{sissa}. In the following we shall show that, if we
subtract the second order self energy of the electrons $\Delta E^0$ , the
contribution to the resulting term is finite and independent from the
normalization volume. This self energy is 
\begin{equation}
\Delta E^0=\sum_{{\bf q}}\nu _q\sum_{{\bf q}^{\prime }{\bf k}^{\prime }}%
\frac{\left| g_{{\bf q^{\prime }},{\bf q}}^{{\bf k}^{\prime }}\right| ^2}{%
\epsilon _q-\epsilon _{q^{\prime }}-\hbar ck^{\prime }}  \label{bb}
\end{equation}
Thus, the {\it observable} contribution to the shift is:

\[
\Delta \widetilde{E}_\omega ^{\left( 2\right) }=\Delta E_\omega ^{\left(
2\right) }-\Delta E^0 
\]

Using (\ref{b4}) and (\ref{bb}), 
\begin{equation}
\Delta \widetilde{E}_\omega ^{\left( 2\right) }=-\sum_{{\bf k}^{\prime
}}\sum_{{\bf qq}^{\prime }}\nu _q\nu _{q^{\prime }}\frac{\left| g_{{\bf %
q^{\prime }},{\bf q}}^{{\bf k}^{\prime }}\right| ^2}{\epsilon _q-\epsilon
_{q^{\prime }}-\hbar ck^{\prime }}  \label{bb6}
\end{equation}

We now show that $\frac 1V{\bf \,}\Delta \widetilde{E}_\omega ^{\left(
2\right) }$ has a finite value, independent from the normalization volume .

Since in low density conditions $\nu _q\nu _{q^{\prime }}\cong z^2\exp
\left[ -\beta \left( \epsilon _{q^{\prime }}+\epsilon _q\right) \right] $,
we have, using the explicit expression (\ref{a1}) for the $g_{{\bf q^{\prime
}},{\bf q}}^{{\bf k}^{\prime }}$, 
\[
\frac 1V{\bf \,}\Delta \widetilde{E}_\omega ^{\left( 2\right) }=z^2\frac{e^2%
}{m^2c}\frac{\hbar ^3}2\frac 1{V^2}\sum_{{\bf k\,q\,q}^{\prime }}\frac{%
\left( {\bf q}^{\prime }\cdot {\bf u}_{{\bf k}}\right) ^2}k\frac{\exp \left(
-\beta \left( \epsilon _{q^{\prime }}+\epsilon _q\right) \right) }{\hbar
ck+\epsilon _{q^{\prime }}-\epsilon _q}\delta _{{\bf k}+{\bf q}^{\prime },%
{\bf q}} 
\]
Summing over ${\bf q}^{\prime }$ and using $\frac 1{V^2}\sum\limits_{{\bf qk}%
}\rightarrow \frac 1{\left( 2\pi \right) ^6}\int d{\bf k}\int d{\bf q}$, we
get

\begin{equation}
\frac 1V\Delta \widetilde{E}_\omega ^{\left( 2\right) }=z^2\frac{e^2}{m^2c}%
\frac{\hbar ^3}{2\left( 2\pi \right) ^6}\int d{\bf q}\int d{\bf k}\frac{%
\left( {\bf q}\cdot {\bf u}_{{\bf k}}\right) ^2}k\frac{\exp \left[ -\beta
\left( \epsilon _{\left| {\bf q}-{\bf k}\right| }+\epsilon _q\right) \right] 
}{\hbar ck+\epsilon _{\left| {\bf q}-{\bf k}\right| }-\epsilon _q}
\label{quelloli}
\end{equation}

The integral over ${\bf k}$ can be computed in spherical coordinates ($%
r,\theta ,\varphi $) choosing ${\bf q}$ as polar axis; the polarization
vector (normal to the polar axis) is defined by $\varphi =0$. We have ($\mu
=\cos \theta $): 
\[
\left( {\bf q}\cdot {\bf u}_{{\bf k}}\right) ^2=q^2\cos ^2\varphi \sin
^2\theta =q^2\cos ^2\varphi \left( 1-\mu ^2\right) 
\]
\[
\epsilon _{\left| {\bf q}-{\bf k}\right| }\pm \epsilon _q=\frac{\hbar ^2}{2m}%
\left( q^2+k^2-2qk\mu \pm q^2\right) 
\]
Calculating the integral (\ref{quelloli}) over $\varphi $, then integrating
over all the possible directions of ${\bf q}$: 
\begin{equation}
\frac 1V\Delta \widetilde{E}_\omega ^{\left( 2\right) }=\frac{z^2e^2}{m^2c^2}%
\frac{\hbar ^2}{2^5\pi ^4}\int_0^{+\infty }dk\int_0^{+\infty
}dqq^4\int_{-1}^1d\mu \left( 1-\mu ^2\right) \frac{\exp \left[ -\left(
2q^2+k^2-2qk\mu \right) /k_T^2\right] }{1+\frac \hbar {mc}\left( \frac k2%
-q\mu \right) }  \label{b41}
\end{equation}
with $k_T:=\sqrt{\frac{2mk_BT}{\hbar ^2}}$ $.$

In order to calculate the integrals over $q$ and $k$ , we notice that the
integrand is not exponentially zero only when the positive definite quantity 
$2q^2+k^2-2qk\mu $ is smaller then a few $k_T^2$'s. This condition is
satisfied only if $k$ and $q$ are small with respect to some $k_T$'s. Then,
if the plasma temperature is not higher than $\sim 10^6\,K$ , we can neglect 
$\frac \hbar {mc}\left( \frac k2-q\mu \right) $ in the denominator of (\ref
{b41}) with respect to $1$ (in fact $\frac \hbar {mc}\left( \frac k2-q\mu
\right) ={\cal O}\left( \frac \hbar {mc}k_T\right) ={\cal O}\left( \sqrt{%
\frac{k_BT}{mc^2}}\right) $, and, if $T\sim 10^6\,K$, $\sqrt{\frac{k_BT}{mc^2%
}}\sim 10^{-2}$).

Now the integrals are Gaussian, and they can be easily evaluated . This
gives 
\[
\frac 1V{\bf \,}\Delta \widetilde{E}_\omega ^{\left( 2\right) }=\frac{z^2e^2%
}{m^2c^2}\frac{\hbar ^2}{2^9\pi ^3}k_T^6 
\]
Finally in the low density assumption ($z\ll 1$) the density $n$ of the
plasma is: 
\begin{equation}
n=z\cdot k_T^3{\frac 1{2^3\pi ^{{\frac 32}}}}  \label{b9}
\end{equation}
As a consequence, we have 
\begin{equation}
\frac 1V{\bf \,}\Delta \widetilde{E}_\omega ^{\left( 2\right) }=\frac{n^2e^2%
}{m^2c^2}\frac{\hbar ^2}{2^3}=\frac{n^2}{m^2c}\frac{\hbar ^3}{2^3}\alpha
\label{b10}
\end{equation}
If the plasma is not homogenous, this energy shift depends on the position
via the electronic plasma thermodynamic parameters. As we noticed in the
introduction, the dependence on the position is negligible over distances of
the order of the wavelength. Thus, the observed energy shift is simply the
volume average (see section \ref{V} below) of $\Delta \widetilde{E}_\omega
^{\left( 2\right) }$: 
\[
\hbar \delta \omega ^{\left( 2\right) }:=\frac 1V\int_{{\cal V}}d{\bf x\,}%
\Delta \widetilde{E}_\omega ^{\left( 2\right) }\left( {\bf x}\right) = 
\]
\begin{equation}
=\frac{\hbar ^3}{2^3m^2c}\alpha \int_{{\cal V}}d{\bf x\,}n^2\left( {\bf x}%
\right)  \label{b11}
\end{equation}

\section{Classical contributions to the shift}

\label{IV}We now consider the contribution to the shift due to $\Delta
E_\omega ^{\left( 1\right) }$. In low density conditions we have, after
summing with respect to ${\bf q}^{\prime }$ and taking the continuous limit,

\[
\Delta E_\omega ^{\left( 1\right) }=\frac{e^2}{m^2}\frac{\hbar ^3}{2\omega }z%
\frac 1{\left( 2\pi \right) ^3}\int d{\bf q}e^{-\beta \epsilon _q}\left( 
{\bf q}\cdot {\bf u}_{{\bf k}}\right) ^2\left[ \frac 1{\hbar \omega
+\epsilon _q-\epsilon _{\left| {\bf q+k}\right| }}-\frac 1{\hbar \omega
+\epsilon _{\left| {\bf q-k}\right| }-\epsilon _q}\right] 
\]
The integral can be calculated in the same approximation that leads to (\ref
{b10}). We obtain 
\begin{equation}
\Delta E_\omega ^{\left( 1\right) }=\frac{e^2}m\frac \hbar {2ck}\frac{k_BT}{%
mc^2}n  \label{c0}
\end{equation}
Once again, if the plasma is not homogeneous, 
\begin{equation}
\hbar \delta \omega ^{\left( 1\right) }:=\frac 1{\hbar V}\int_Vd{\bf x\,}%
\Delta E_\omega ^{\left( 1\right) }\left( {\bf x}\right) =\frac{e^2}m\frac 1{%
2ck}\frac{k_BT}{mc^2}\frac NV  \label{c9}
\end{equation}
where $N$ is the total number of electrons in $V$.

Let us now determine the contribution coming from $H_I^{^{\prime \prime }}$
of (\ref{a0}).

Using again a perturbative treatment we see that the first order term is no
longer zero and its magnitude in terms of powers of the coupling parameter $%
\alpha $ is the same as that of (\ref{b11}), it cannot then be a priori
overlooked.

The correction we are now studying is: 
\begin{equation}
\hbar \delta \omega ^{\prime \prime }=\sum_{{\bf q}}\nu _q\left\langle \Phi
_\omega \otimes \Psi _{{\bf q}}\left| \widehat{H}_I^{\prime \prime }\right|
\Phi _\omega \otimes \Psi _{{\bf q}}\right\rangle  \label{c1}
\end{equation}
where $\widehat{H}_I^{\prime \prime }$ is given by eq. (\ref{a6}). Thus 
\[
\hbar \delta \omega ^{\prime \prime }=\frac{e^2}{2mc^2}\frac \hbar V\sum_{%
{\bf q}^{\prime }}\nu _{q^{\prime }}\left\langle \Phi _\omega \otimes \Psi _{%
{\bf q}}\left| \sum_{{\bf q},{\bf q}^{^{\prime }},{\bf k,k}^{\prime }}\frac{%
\delta _{{\bf q-q}^{\prime },{\bf k-k}^{\prime }}}{\sqrt{kk^{\prime }}}b_{%
{\bf q}^{\prime }}^{\dagger }b_{{\bf q}}\otimes a_{{\bf k}^{\prime
}}^{\dagger }a_{{\bf k}}\right| \Phi _\omega \otimes \Psi _{{\bf q}%
}\right\rangle = 
\]
\begin{equation}
=\frac{e^2}{2mc^2}\frac \hbar V\frac 1\omega \sum_{{\bf q}^{\prime }}\nu
_{q^{\prime }}=\frac{e^2}{2mc}\frac \hbar k\frac NV  \label{c2}
\end{equation}

In fact $\sum_{{\bf q}^{\prime }}v_{q^{\prime }}$ is, by definition, the
total number $N$ of particles, consequently $\frac NV$ is nothing else than
the average density $\left\langle n\right\rangle $ of the plasma in the
given volume. This gives $\delta \omega ^{\prime \prime }=e^2\left\langle
n\right\rangle /(2mck)$, which is precisely the classical correction to the
dispersion relation due to the presence of a low density homogeneous
electron plasma (whose constant density is $\left\langle n\right\rangle $),
when the temperature is zero. In fact, when it is $\frac{\omega ^2}{c^2k^2}>>%
\frac{k_BT}{mc^2}$, we have (see \cite{plasma1}) 
\begin{equation}
\omega ^2=\omega _P^2+c^2k^2\left( 1+\frac{k_BT}{mc^2}\frac{\omega _P^2}{%
\omega ^2}\right)  \label{c3}
\end{equation}
where $\omega _p=\sqrt{\frac{e^2}mn}$ is the plasma frequency. For $\omega
_p\ll kc$ and $\frac{k_BT}{mc^2}\ll 1$ ( non relativistic approximation)
this gives 
\begin{equation}
\omega =kc+\frac 12\left( 1+\frac{k_BT}{mc^2}\right) \frac{\omega _p^2}{kc}%
=kc+\left( 1+\frac{k_BT}{mc^2}\right) \frac{e^2n}{2mck}  \label{c4}
\end{equation}

This is of course true when $T\sim 10^6$ K, which implies $\frac{k_BT}{mc^2}%
\sim 10^{-4}$.

The last term in (\ref{c4}) coincides with $\delta \omega ^{\prime \prime
}+\delta \omega ^{\left( 1\right) }$, where $\delta \omega ^{\left( 1\right)
}$ gives the first order temperature correction. Like all the classical
terms, it depends on $N/V$ ($V$ is the normalization volume of the
electromagnetic field). Hence, when $V\rightarrow \infty $, such a term
gives a vanishing contribution to the frequency shift.

\section{Discussion}

\label{V}

The dependency of the classical terms on $N/V$ rules them out when the
measuring apparatus is outside of the plasma and the normalization volume
for the electromagnetic wave is infinite. {\em This is not the case for the
quantum contribution }(\ref{b11}), which is always different from zero: it
will be the only contribution to the frequency shift observable in
astrophysical or even cosmological conditions. An important remark on (\ref
{b11}) is that the volume over which one integrates cannot be the volume of
the whole universe; in fact it extends from the source to the receiver ($%
\sim $ from $-\infty $ to $+\infty $ ) along the line of sight, but
transversely it should not be more than the distance over which the plasma
may practically be thought of as infinite for quantum mechanical
calculations. This transversal extension is a sort of coherence length for
the plasma. It should be less than the wavelength of plasmons, the Debye
length and the screening length of the plasma. On the other hand its square
should of course be much greater than the Compton scattering cross section:
our approach has indeed nothing to do with individual scattering phenomena.

Assuming for simplicity a Gaussian distribution along the line of sight,
such as 
\[
n\left( x\right) =n_0e^{-\frac{x^2}{2R^2}} 
\]

(\ref{b11}) gives: 
\begin{equation}
\delta \omega ^{\left( 2\right) }=\frac{\hbar ^2}{2^3m^2c}\alpha n_0^2\pi
^{3/2}RL^2  \label{b12}
\end{equation}

$L$ is the transverse ''coherence length'': it may in fact be used as a
phenomenological parameter.

In our model relaxation phenomena in the plasma play no r$\widehat{\text{o}}$%
le. This is because we consider a steady state situation and, furthermore,
our plasma is modelled as a reservoir, that can exchange with the photon any
amount of energy without changing its thermodynamic state. Thus any dynamic
process due to the transit of the photon is neglected. Of course, this
approximation is good only if the plasma is very extended, its density is
very low and the electromagnetic field is not too strong, but all of these
conditions are indeed satisfied in our case.

The result we found has been obtained under some assumptions that need a
careful consideration. First of all, the calculation is based on the
perturbation theory. Usually this theory treats single electrons interacting
with an external field or with a bath of photons. Here we are dealing with
one photon interacting with a gas of electrons. The fermionic nature of
electrons comes into play through the factor $\nu _{{\bf q}}\nu _{{\bf q}%
^{\prime }}$ in (\ref{bb6}), or simply, when the approximation is
appropriate, through the fugacity $z$ of the plasma. When the fugacity is
small enough, it is easy to see that the relevant terms of the perturbative
series are weighed not only by $\alpha $ and its powers, but rather by
products of powers of $\alpha $ and powers of $z$: this fact may change the
relative importance of the different contributions.

Actually (\ref{b11}) is of order $\alpha z^2$. If relativistic effects had
been taken into account, a second order contribution had come also from
processes such as virtual pair creation. Such a contribution is proportional
to $\alpha z\left( \frac{\hbar \omega }{mc^2}\right) ^2$ and can be
overlooked when 
\[
\left( \frac{\hbar \omega }{mc^2}\right) ^2<<z 
\]

To sum up we conclude that the blue shift found in this paper applies to
waves whose frequency satisfies to the condition 
\[
\omega _p<\omega <<\frac{mc^2}\hbar \sqrt{z} 
\]
For higher frequencies relativistic contributions must be included.

To give an idea of the numbers, consider that in the physical conditions of
vast portions of the solar corona $z$ can be as low as $10^{-17}$;
consequently the upper frequency is $\sim 10^{11}$ Hz.


\begin{references}
\bibitem{ginzburg1}  V.L.Ginzburg, {\it Propagation of electromagnetic Waves
in Plasma}, Gordon and Breach Pub., New York (1961)

\bibitem{wolf1}  J. T. Foley, E. Wolf, Phys. Rev. A,{\bf \ 40, }579, (1989)

\bibitem{wolf2}  J. T. Foley, E. Wolf, Phys. Rev. A,{\bf \ 40, }588, (1989)

\bibitem{wolf3}  D. F. V. James, E. Wolf, Phys. Lett. A, {\bf \ 146, }167,
(1990)

\bibitem{Landau}  L. Landau, E. Lifshitz, M\'{e}canique Quantique, 510, MIR,
Moscow (1967)

\bibitem{sakurai}  H. Haken, Quantum field theory of solids, North Holland,
Amsterdam (1976), section 44b)

\bibitem{sissa}  L. Accardi, A. Laio, F. Cardi, G. Rizzi, in Proc. 11th
Italian conf. on General Relativity, Trieste, 1994

\bibitem{plasma1}  Page 119 of ref.\cite{ginzburg1}
\end{references}
\end{document}